\documentclass[aps,pra,twocolumn,twocolumn,superscriptaddress]{revtex4-1} %maybe a better one?
\usepackage{amsmath}
\usepackage{latexsym}
\usepackage{graphicx}
\usepackage{amsfonts}
\usepackage{braket}
\usepackage{natbib}
\usepackage{amssymb}
\usepackage{fancyhdr}
\usepackage[utf8]{inputenc}
\usepackage{dsfont}
\usepackage{colortbl}
\usepackage{xcolor}
\usepackage{subfigure}
\usepackage{psfrag}
\usepackage[colorlinks=true,citecolor=blue,linkcolor=blue]{hyperref}
\usepackage{tikz}
\usepackage{ulem}
\usepackage{bbold}
\usepackage{bbm}
\allowdisplaybreaks
\newcommand{\ope}[1]{{\ensuremath{\hat{\text{#1}}}}}
\newcommand\bbone{\ensuremath{\mathbbm{1}}}
\usetikzlibrary{arrows,shapes}

\newcommand{\oold}[1]{\emph{\color[rgb]{.4,.4,.4}(old version see tex file)}}
\newcommand{\ooold}[1]{\emph{\color[rgb]{.4,.4,.4}(ovtf)}}

\renewcommand{\section}[1]{\emph{#1} -- }

\newcommand{\PTrace}[2]{\text{Tr}_{#1}\left[#2\right]}	% Partial Trace

\newcommand{\ketbra}[2]{\ket{#1}\hspace*{-0.12cm}\bra{#2}}

\renewcommand{\cal}[1]{\ensuremath{\mathcal{#1}}}

\begin{document}

\title{Benchmarking non-simulable quantum processes via symmetry conservation}
\author{Tobias Chasseur}
\affiliation{Theoretical Physics, Saarland University, 66123 Saarbr\"ucken, Germany}
\author{Felix Motzoi}
\affiliation{Theoretical Physics, Saarland University, 66123 Saarbr\"ucken, Germany}
\affiliation{Department of Physics and Astronomy, Aarhus University, 8000 Aarhus C, Denmark}
\author{Michael Kaicher}
\affiliation{Theoretical Physics, Saarland University, 66123 Saarbr\"ucken, Germany}
\author{Pierre-Luc Dallaire-Demers}
\affiliation{Theoretical Physics, Saarland University, 66123 Saarbr\"ucken, Germany}
\affiliation{Department of Chemistry and Chemical Biology, Harvard University, Cambridge, MA 02138 USA}
\author{Frank K. Wilhelm}
\affiliation{Theoretical Physics, Saarland University, 66123 Saarbr\"ucken, Germany}

\begin{abstract}
As quantum devices scale up, many-body quantum gates and algorithms begin to surpass what is possible to simulate classically. Validation methods which rely on such classical simulation, such as process tomography and randomized benchmarking, cannot efficiently check correctness of most of the processes involved. In particular non-Clifford gates are a requirement for not only universal quantum computation but for any algorithm or quantum simulation that yields fundamental speedup in comparison with its classical counterpart.  We show that it is in fact still possible to validate such non-simulable processes by taking advantage of expected or engineered conservations laws in the system, combined with a unitary one-design strategy to randomize errors over the computational Hilbert space.  We show that in the context of (fault-tolerant) quantum error correction, we can construct a one-design using the logically encoded Clifford group over the engineered error-free stabilizer subspace to obtain average error for arbitrary logically-encoded gates and algorithms.  In the case of benchmarking simulation of physical systems, these can have various exotic symmetries over which one-design strategies can nonetheless be constructed.  We give examples for fermionic systems which conserve particle number, as well as for the Fermi-Hubbard model. The symmetry benchmarking method preserves robustness to state preparation and measurement imperfections.
\end{abstract}
\maketitle

A requirement for the successful development of quantum technologies, in addition to designing performance enhancing protocols, and building hardware on which they can run, is being able to ascertain their high-fidelity operation.

This subfield of quantum information broadly groups together the verification, benchmarking, or characterization of underlying black-box physical processes.  Which metric can be evaluated typically depends on its complexity.  For example, verification of cryptographic black-box security can often be demonstrated via a Bell test \cite{bell1964einstein,aspect1982experimental}, while potential solutions to NP problems can be verified classically in polynomial time \cite{goldreich2010p}.  On the other hand, full characterization of an unknown process through process tomography scales exponentially with system size and thus is tractable only for small dimension or sparse Liouvillians \cite{gross2010quantum,cramer2010efficient,gross2011recovering,shabani2011efficient}.  A third branch of validation methods has been developed known as Randomized Benchmarking (RB) \cite{Knill2008,Magesan2011}, which scales polynomially with system size and amplifies deviations from the ideal process with respect to other sources of error such as preparation and measurement.

However, complex dynamical protocols such as digital quantum computation \cite{Nielsen2000}, adiabatic quantum computation \cite{farhi2001quantum}, and quantum simulation \cite{lloyd1996universal,aspuru2005simulated,Kassal2011,mcclean2016theory,Barends2015,Wecker2015a,Bauer2016,DallaireDemers2016b,Salathe2015,Martinez2016,Zohar2017} are typically useful because classical emulation of the same tasks can require significantly more time. 
Yet most processes that can be validated to date involve only classically simulable ideal outcomes.  As such, earlier proposals to expand the purview of RB, to include benchmarking individual operations \cite{Magesan2012}, to remove assumptions about leakage and gate-dependent errors \cite{Chasseur2015}, and to test certain non-Clifford gates using a different basis \cite{Wallman2014X,Carignan-Dougas2015,Cross2016}, are nonetheless restricted to processes that have efficient equivalent classical circuits. The benchmarking of arbitrary evolution on the other hand, has shown to result in exponential scaling \cite{Chasseur2016,Reich2013}.  

In this work, we present a method to efficiently verify symmetry  conservation laws in sequences of arbitrary quantum operations. We achieve this by drawing a distinction between randomizing input states to the unknown noise afflicting a finite group of operations (which allows diffusion of errors into a single average error channel), and the random output state of a sequence of quantum operations due to the inherent complexity of a large random Hilbert space.  The latter can nonetheless exhibit structure 
in the form of either inherent stabilized subgroups (due to the form of the dynamics) or engineered redundancies (due to large Hamming distances in encoded subspaces). These conservation laws can of course vary between applications, but typically be found anywhere from the algorithmic level down to the hardware implementation level. Mathematically, to this end, we generalize the benchmarking requirements to eliminate the use of a so-called `2-design', for which the randomization over the finite group must be classically tracked and subsequently inverted (hence the process is `squared').  Instead, by constructing a '1-design' strategy, we randomize not only the process itself, but also the desired final outcome within given boundaries. We derive a suitable generalized metric to assess error propagation outside of any problem-specific conserved subspace. 

We conclude with some examples of typical conserved computational spaces.  As the simplest application of the algorithm, at the lowest level of hardware architecture, we often see computational spaces whose population is preserved (e.g.~avoiding auxiliary or leakage subspaces \cite{Chasseur2015}). More generally, local quantum gate operations will have some number of invariants, which they conserve by virtue of being typically generated by low rank operators.  Many-body quantum simulations of more exotic physical system also offer many examples of conserved symmetries. In this work we give in particular prescriptions for how to use number conservation and parity conservation to benchmark population degradation. Finally, at the highest algorithmic level, quantum annealing, cryptography, and fault-tolerant computation all use enforced redundant information to improve system performance.  We detail how to use stabilizer-code conservation to benchmark average error propagation in such encoded systems.

\section{Symmetry Benchmarking Protocol}
We wish to assess a given (set of) error channel(s) that take us out of a restricted subspace $\mathcal{H}_0$.  The ideal dynamics of the system preserve the eigenstates of a conserved operator \ope{C}, i.e. a stabilizer of the system. Let $\lambda_\gamma$ be the degenerate eigenvalues of   \ope{C}. To conserve the symmetry, all operations in the algorithm (gates) must be block--diagonal in the \ope{C}--eigenbasis, with the blocks corresponding to the eigenspaces. 
%A prominent example we will use later in this work is the particle number in molecules which is constant in quantum chemistry simmulations. 
The approach of the proposed protocol is to find a set $\mathcal{D}$ of gates on $\mathcal{H}=\bigoplus_\gamma \mathcal{H}_\gamma$ that acts as a unitary one--design on any of the eigenspaces $\mathcal{H}_\gamma$. A unitary one--design is defined by having the same probability distribution as the Haar--measured special unitary group $\mathcal{SU}$ for first order polynomial functions in any gate and its adjoint. In particular this means
\begin{align}
 \frac{1}{\sharp\mathcal{D}}\sum_{\ope{D}\in\mathcal{D}} \ope{D}\hat{\rho}\ope{D}^\dagger=\int_\mathcal{SU}\text{d}\ope{U}~\ope{U}\hat{\rho}\ope{U}^\dagger,\label{eqn:od}
 \end{align}
with $\sharp$ denoting the cardinality. We define the symmetry breaking $\mu$ as the average population decay out of an initial \ope{C}--eigenspace caused by an error channel $\Lambda$.  We estimate the symmetry preservation $\Gamma=1-\mu$ via a RB-like protocol applying random one--design sequences of different lengths $y$ and measuring the population of the initially populated subspace $\mathcal{H}_{\gamma_0}$.  Using that $\mathcal{D}$ is a unitary one--design one obtains the average symmetry preservation of a sequence as  
\begin{align}
\Gamma_y &\equiv\frac{1}{\sharp\mathcal{D}^y}\sum_{\{D_j\}\in\mathcal{D}^y}\PTrace{\gamma_0}{\left(\prod_{j=y }^1(\Lambda D_j)\right)(\hat{\rho}_0)}\label{eqn:asb}\text{.}
\end{align}
Here $\PTrace{\gamma_0}{~}$ denotes the trace over the preserved subspace $\mathcal{H}_{\gamma_0}$, the unhatted gates describe the effect on density matrices (superoperators) and the inverse order of the product ensures the correct succession of the quantum gates. We make use of the following definition: the \emph{half twirl} of $\Lambda$ over \cal{D} is $\Lambda_{\rm ht}\equiv \frac{1}{\sharp\mathcal{D}}\sum_{D\in\mathcal{D}}\Lambda D$, in contrast to the usual twirl $\Lambda_{\rm twirl}=\frac{1}{\sharp\mathcal{C}}\sum_{C\in\mathcal{C}}C\Lambda C^{-1}$  over a group $\mathcal{C}$ \cite{Dankert2009}. Similarly to the arguments in \cite{Chasseur2015}, $\Lambda_{ht}$ to the power of $y$ is acted on by a linear functional, hence it can be simplified to
\begin{align}
\Gamma_y &=\vphantom{\sum_j^j}\PTrace{\gamma_0}{\Lambda_{\rm ht}^y (\hat{\rho}_0)}=\sum_i \alpha_i \lambda_i^y
\end{align}
As $\Lambda_{\rm ht}$ is a completely positive, trace preserving map, the entries on its matrix representation are real and positive and hence the absolute values of the $\lambda_i$ are smaller than or equal to one due to the Perron--Frobenius theorem \cite{Perron1907,Frobenius1912,Meyer2000}. This implies that the population decay can be fitted with just a few exponential decays despite the maximum number of different eigenvalues scaling as $d^2\equiv2^{2n}$ \cite{Chasseur2015}. Finally, we can extract the averaged error $\Lambda$ per time step as
\begin{align}
\mu&=1-\int_{\mathcal{SU}(d_0)}\PTrace{\gamma_0}{\Lambda\left(\ope{U}\hat{\rho}_0\ope{U}^\dagger\right)}~\text{d}\ope{U}\\&=1-\frac{1}{\sharp\mathcal{D}}\sum_{D\in\mathcal{D}}\PTrace{\gamma_0}{\left(\Lambda D\right)(\hat{\rho}_0)}=1-\Gamma_1
\text{,}
\end{align}
namely, the symmetry breaking of $\ope{C}$.  The protocol inherits robustness against state preperation and mesurement (SPAM) errors, similarly to Clifford benchmarking protocols \cite{Knill2008,Magesan2011,Chasseur2015}. This stabilizer leakage quantifies the error accumulation for any error channel that causes decays out of it.  When all error channels are predominantly manifested via decay out of the conserved subspace (i.e. the Hamming distance of the stabilized symmetry is large), this gives a metric for the cumulative average Haar-measure error.  We will give examples for both cases.

\section{Benchmarking arbitrary operations}
The error randomization over the one-design allows us to also benchmark operations outside of the set \cal{D}.  Thus, we introduce a second set of operations that we want benchmark with respect to the error channel, which we call $\mathcal{I}$, containing one, several or all possible gates of the algorithm. Inspired by Interleaved Randomized Benchmarking (IRB) \cite{Magesan2012} we interleave the random \cal{D}--sequence with random elements of \cal{I} to assess a combined stabilizer decay $\mu_\cal{ID}$. The symmetry preservation for that combined sequence of length $2y$ gives
\begin{align}
\Gamma_y &=\frac{1}{\sharp\mathcal{D}^y\sharp\cal{I}^y}\sum_{\{I_j\},\{D_j\}}\PTrace{\gamma_0}{\left(\prod_{j=y}^1(I_j\Lambda_\cal{I}\Lambda_\cal{D} D_j)\right)(\hat{\rho}_0)}\\
&=\frac{1}{\sharp\mathcal{D}^y\sharp\cal{I}^y}\sum_{\{I_j\},\{D_j\}}\PTrace{\gamma_0}{\left(\prod_{j=y}^1(\Lambda_\cal{ID} D_j)\right)(\hat{\rho}_0)}\text{,}
\end{align}
where $\Lambda_{\mathcal{ID}}\equiv \Lambda_{\mathcal I}\Lambda_{\mathcal D}$.
As before we can derive $\Gamma_1$, or $\frac{\Gamma_1}{\Gamma_0}$ respectively, to assess the combined error $\mu_{\mathcal{ID}}$. An estimate for the average decay rate of \cal{I} is given by $\mu_\mathcal{I}\approx\mu_\cal{ID}-\mu_\cal{D}$; this as well as strict bounds are derived as for Interleaved Randomized Benchmarking. However, in this case, we can also efficiently benchmark errors for operations that cannot be simulated classically, such as quantum algorithms themselves or non-Clifford gates.  The latter are needed for universal quantum computation and are thus required for full verification of a quantum computer.

\section{Quantum chemistry: number conservation}
A prominent symmetry in quantum simulation of physical systems, as well as many gate architectures for quantum computing (e.g.~iSWAP interactions) is the conservation of excitation or particle number. Most commonly, this symmetry arises when mapping from the para-fermionic to the fermionic basis \cite{jordan1993paulische,whitfield2011simulation,bravyi2002fermionic,Tranter2015}. The electron number operator $\ope{n}\equiv\sum_i \ope{n}_i\equiv\ope{C}$ divides the Hilbert space into $n+1$ eigenspaces $\mathcal{H}_\gamma$ with $0\leq \gamma\leq n$ excited qubits and dimension ${n \choose \gamma}$. 
%However, it is not a one--design on all subspaces independently and  simultaneously, i.e., ${\mathcal{D}\neq\prod_\gamma \bbone_0\oplus\cdots\mathcal{D}_{\gamma}\cdots\oplus\bbone_{n}}$. We discuss how that would be useful for further simplifications and which properties of such a simultaneous one--design are given for our particular solution later on.

To properly define the conditions for $\mathcal{D}$ being a one--design we have to review and define a basis for the Hilbert space $\mathcal{R}_{\gamma_0}$ of density operators of states in $\mathcal{H}_{\gamma_0}$. %\mkcomment{define a basis for the Hilbert space $\mathcal{H}_{\gamma_0}$ of density operators of states $\mathcal{R}_{\gamma_0}$}.
$\mathcal{R}_{\gamma_0}$ can be seen as the union of $\left\{\ketbra{i}{i}\right\}_{\ket{i}\in\mathcal{H}_{\gamma_0}}$, $\left\{\ketbra{i}{j}+\ketbra{j}{i}\right\}_{\ket{i},\ket{j}\in\mathcal{H}_{\gamma_0},i<j}$ and $\left\{-i\ketbra{i}{j}+i\ketbra{j}{i}\right\}_{\ket{i},\ket{j}\in\mathcal{H}_{\gamma_0},i<j}$ which we denote $\left\{\ket{\text{B}_i}\right\}$, $\left\{\ket{\text{X}_{ij}}\right\}$ and $\left\{\ket{\text{Y}_{ij}}\right\}$ respectively. Let  $\mathcal{D}$ be the one--design acting on $\mathcal{R}_{\gamma_0}$, then $\sum D_{d\in\mathcal{D}}$ maps any density matrix onto the completely mixed state. The action of the one-design in this basis can then be simply rewritten as
\begin{align}
 \frac{1}{\sharp\mathcal{D}}\sum_{D\in\mathcal{D}} D \ket{\text{B}_i}&=\frac{1}{d_0}\sum_{k}\ket{\text{B}_k}\tag{§1}\\
 \sum_{D\in\mathcal{D}} D \ket{\text{X}_{ij}}&=0\tag{§2}\\
 \sum_{D\in\mathcal{D}} D \ket{\text{Y}_{ij}}&=0\text{,}\tag{§3}
\end{align}
with dim$(\mathcal{H}_{\gamma_0})=d_0$. Because of the linearity of Eq. \ref{eqn:od}, these are the only nontrivial conditions needed for constructing a unitary one--design. Focusing on (§1), we want to ensure that by sampling over $\mathcal{D}$ the transition between each two basis states is realized with equal probability. We implement this using arbitrary qubit permutations to randomly redistribute the excited qubits' sites.  We populate each basis state with equal probability. Averaging over the one--design, this yields the completely mixed state of $\mathcal{H}_{\gamma_0}$ regardless of the initial state, thus satisfying (§1). Note that although qubit permutations implement all transitions between two states it is not equivalent to state permutation. This is in fact crucial for the scalability of this solution and shows importance in the following section.

\begin{figure}[htbp!]
 \centering
 \includegraphics[width=0.45\textwidth]{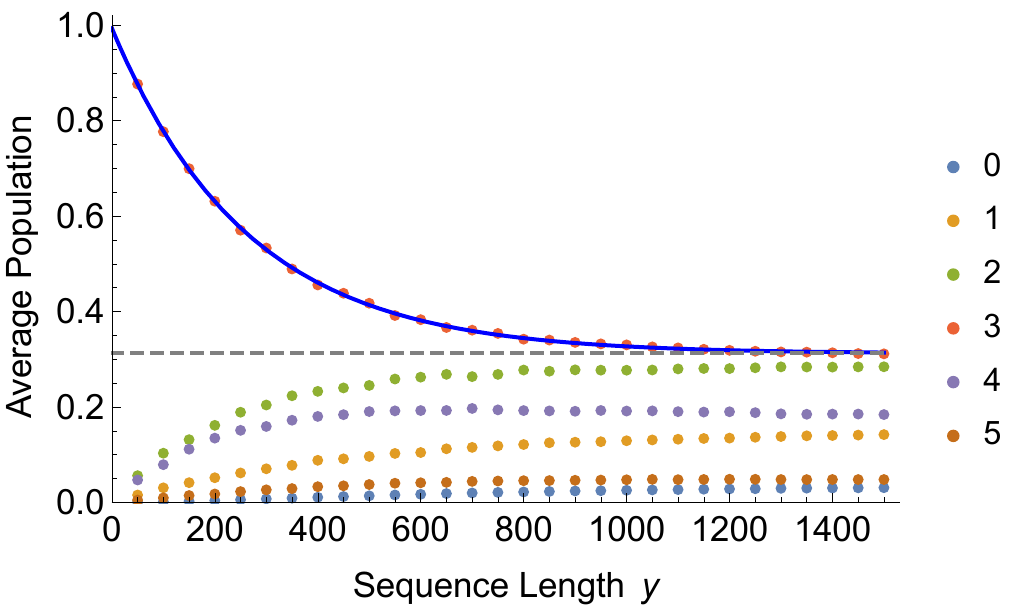}
 \caption{Benchmarking the number preservation symmetry on five simulated qubits. %Every ISWAP includes an error on the two qubits in question which was derived as an arbitrary error channel. The $\gamma=3$ subspace is benchmarked via fit (solid blue line) to have an average population leakage of $\mu=.88\%$ and $\Gamma_1=99.12$.
 }
 \label{fig:int}
\end{figure}

To satisfy the remaining conditions (§2) and (§3), we examine the effect of the qubit permutations via iSWAPs on the X$_{ij}$ and Y$_{ij}$ matrix elements, where each element again is mapped onto a density operator X$_{ij}$, or Y$_{ij}$ on the space $\mathcal{R}_{\gamma_0}$ correspending to the preserved number of excited qubits. Introducing a uniformly random $\pm 1$ phase between every two states ensures that those occur with opposite signs equally likely, hence sum up to zero, satisfying (§2) and (§3). This random phase is not inherently given by the phases included in the iSWAPs but easily achieved by a probability $1/2$ $\hat{\sigma}_Z$--gate
on every qubit. This matches our intuition that X$_{ij}$ and Y$_{ij}$ represent coherent phases between states so that randomizing all phases should eliminate them. The above protocol using the derived unitary one--design is simulated in Fig. \ref{fig:int} on a five qubit system which is initialized in a state which has three excited qubits. Every permutation of qubits consists of iSWAPs which,  in this example, contain predefined errors on the pair of qubits. The error channel is derived as a unitary operator close to the identity acting on a four-qubit Hilbert space, then tracing out two qubits. The $\gamma=3$ subspace is benchmarked via a fit of the population decay to have an average population leakage of $\mu=.88\%$ and $\Gamma_1=99.12$. 

% As stated above $\mathcal{D}$ is a one--design on each subspace individually but not on all subspaces simultaneously, hence it does not necessarily act in the same way on coherence between subspaces, i.e., on X$_{ij}$ and Y$_{ij}$ with states $i,j$ not in the same $\mathcal{H}_\gamma$. For the scope of this work it is sufficient to evaluate the difference in the application of $\sum_{D\in\mathcal{D}}D$ \- one can show that
% \begin{alignat}{4}
% 0 &=\sum_{D\in\mathcal{D}}D \ket{\text{X}_{ij}} &&= \sum_{D\in\mathcal{D}}D \ket{\text{Y}_{ij}},%\notag%&=\sum_{D\in\mathcal{D}_{\rm sim}}D \ket{\text{X}_{ij}} &&= \sum_{D\in\mathcal{D}_{\rm sim}}D \ket{\text{Y}_{ij}}\text{.} 
% \end{alignat}
% i.e., all phase relations between subspaces vanish.

The dynamics of the random sequence can be viewed even more simply. Since applying $\sum_{D\in\mathcal{D}}D$ twice is equivalent to a single application, reviewing the average symmetry preservation of equation (\ref{eqn:asb}) gives 
\begin{align}
\Gamma_y &=\frac{1}{\sharp\mathcal{D}^{2y}}\sum_{\{C_j\}\in\mathcal{D}^y}\sum_{\{D_j\}\in\mathcal{D}^y}\PTrace{\gamma_0}{\left(\prod_{j=y}^1(C_j\Lambda D_j)\right)(\hat{\rho}_0)}\text{.}
\end{align}
The updated $\Lambda_{\rm ht}^\prime=\frac{1}{{\sharp\mathcal{D}}^2}\sum_{C,D}C\Lambda D$ commutes with any unitary evolution within subspaces $\mathcal{H}_\gamma$ and can therefore be reduced to simple transition rates between those subspaces. This not only provides a more easily approachable concept but gives the intuition for the Ansätze in the following sections.

\section{Fermi--Hubbard model and parity conservation}
A symmetry that is often encountered in quantum technologies, such as for measurement-based entanglement generation and for error correction is parity-preserving operations.  In the context of quantum simulation, it appears in the Fermi--Hubbard model, used to study strongly correlated electrons in condensed matter physics including basic atomic structure and second quantization \cite{Lafarge1993}. The computationally most costly parts of its simulations can in principle be resolved  by medium sized quantum computers \cite{Dallaire-Demers2016}. While some of the Hamiltonians employed in that scheme are number conserving and can be treated using the set \cal{D} derived previously, others are not, namely, the terms which induce superconductivity to the model. However, these terms always change the electron number by two (a Cooper pair), preserving parity. 

\begin{figure}[htbp!]
 \centering
 \includegraphics[width=0.45\textwidth]{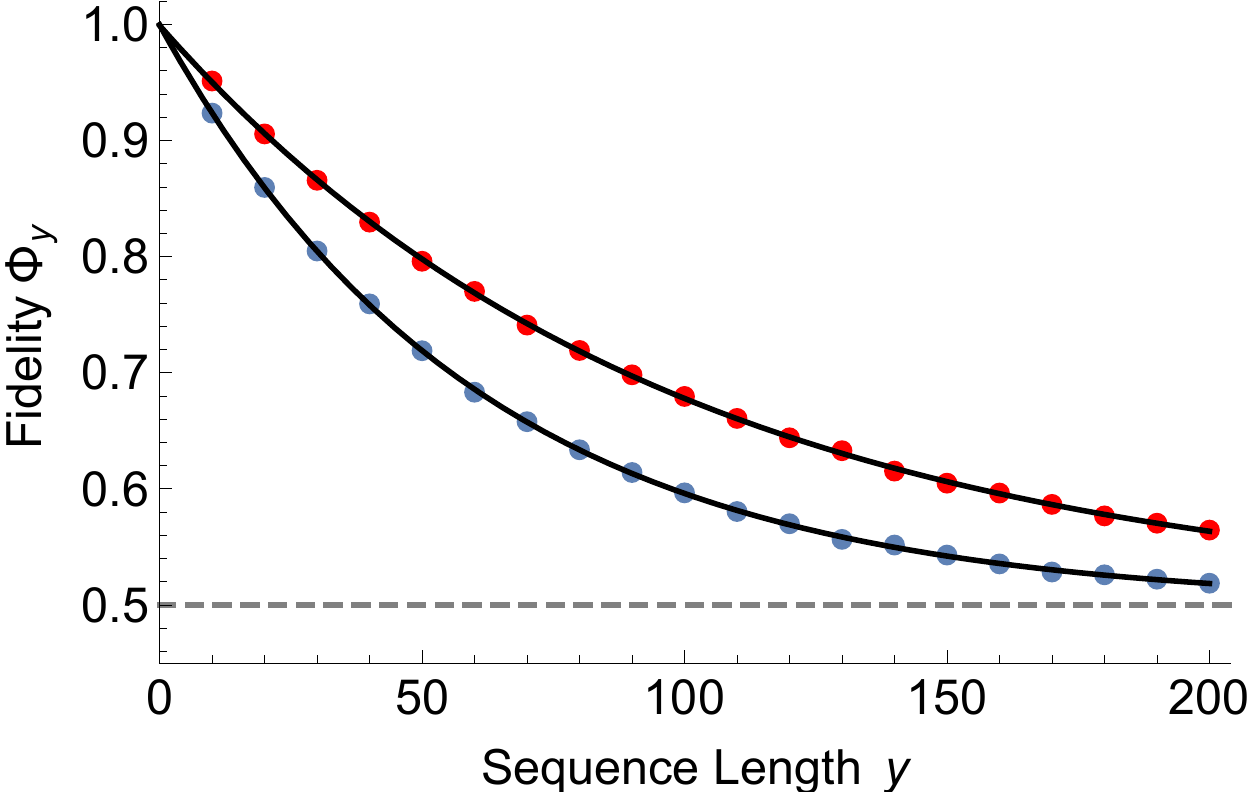}
 \caption{Interleaved Symmetry Benchmarking for the Fermi--Hubbard model on the 6 qubit even subspace for interleaved gate $I=\sigma_2\sigma_3+\sigma_2^\dagger\sigma_3^\dagger$% with an exact symmetrybreaking of $\mu_I=.30\%$. Extracting $\mu_{\rm ISB}=.86\%$ and $\mu_{\mathcal{D}}=.51\%$ provides an estimate of $\mu_I=.35\%$/$\Gamma_1=0.9965$ which is remarkably close to the actual values as $\Lambda_I$ is not the dominating error term.
 }
 \label{fig:isb}
\end{figure}

As the Fermi--Hubbard model involves an even number $n$ of electron sites/qubits, the subspaces $\mathcal{H}_{even \rm}$ and $\mathcal{H}_{odd \rm}$ are of equal dimension $2^{n-1}$; it is in principle possible to map the $n-1$ qubit Clifford group onto those subspaces, but such a protocol would map single qubit gates into multiqubit ones and visa versa, yielding a potentially exponential increase in gate complexity. Instead, we refrain from finding a new unitary one--designs for $\mathcal{H}_{even \rm}$ and $\mathcal{H}_{odd \rm}$ but rely on the transition rates derived previously.
The symmetry preservation on the even subspace for only one individual gate $I$ is
\begin{align}
\Gamma_1 &=\frac{1}{\sharp\mathcal{D}}\sum_{D\in\mathcal{D}_{\rm even}}\PTrace{\rm even}{(I\Lambda_\cal{I}\Lambda_\cal{D} D)(\hat{\rho}_0)}\\
&= \frac{1}{2^{n-1}}\PTrace{\rm even}{(I\Lambda_\cal{I}\Lambda_\cal{D})(\bbone_{\rm even})}\\
\intertext{as a unitary one--design on the even subspace would map to the identity thereon. Writing it as sum of the identities of different subspaces gives}
%&=\sum_{\gamma~\text{even}} \frac{1}{2^{n-1}}\PTrace{\rm even}{(I\Lambda_\cal{I}\Lambda_\cal{D})(\bbone_{\gamma})}\\
&=\sum_{\gamma~\text{even}} \frac{d_\gamma}{2^{n-1}}\sum_{D\in\mathcal{D}}\PTrace{\rm even}{(I\Lambda_\cal{I}\Lambda_\cal{D} D)(\hat{\rho}_{\gamma})}\\
&\equiv \sum_{\gamma~\text{even}} \frac{d_\gamma}{2^{n-1}} \Gamma_1^{\gamma},
\end{align}
where $\hat{\rho}_{\gamma}$ is an initial state in the respective subspace. Each of those $\Gamma_1^{\gamma}$ can be derived by sequences of the usual form; an estimation for $\mu_\mathcal{I}$ can be obtained for each subspace via interleaved symmetry benchmarking and subsequently an overall estimation can be found. As there are only $\frac{n}{2}$ or $\frac{n}{2}+1$ different subspaces, this scales linearly in $n$ and is therefore efficiently scalable in the number of qubits. The protocol also translates easily to a set \cal{I} of gates allowing for an efficient symmetry benchmarking; Figure \ref{fig:isb} shows the data of a simulation for six qubits where we benchmark the symmetry preservation on the even subspace for a interleaved gate $I=\sigma_2\sigma_3+\sigma_2^\dagger\sigma_3^\dagger$ with an exact symmetry breaking of $\mu_I=.30\%$.  Extracting $\mu_{\rm ISB}=.86\%$ and $\mu_{\mathcal{D}}=.51\%$ provides an estimate of $\mu_I=.35\%$/$\Gamma_1=0.9965$ which is remarkably close to the actual values as $\Lambda_I$ is not the dominating error term. This better-than-expected performance is similar to observations from Interleaved Randomized Benchmarking \cite{Magesan2012}.

\section{Benchmarking logically encoded processes}
This protocol allows benchmarking not only inherent symmetries in (simulated) physical systems, but engineered symmetries for which the vast majority (asymptotically speaking) of error leaks through particular `syndrome' states, that is, avoiding direct transitions between logical stabilizer eigenstates.  In this case, we can expect to not only quantify important sources of error, but all sources of error present.  There are many examples of engineered redundancy for the purposes of error suppression or monitoring, most notably in (fault-tolerant) quantum error correcting codes (ECC), but also in quantum cryptography, simulation, and adiabatic quantum computation.

Here we are interested in showing that we can obtain error metrics that are amplified and immune to SPAM, as in standard RB, but also that it allows us to do so for entire algorithms and arbitrary logical gates, such as non-Clifford gates which are fundamental to the logically-encoded quantum computation.

We consider here a Hilbert space encoded in the form $\mathcal{H}=\{0,1\}^{\otimes n}\otimes\{0,1\}^{\otimes m}$ where the latter space is used for syndrome measurements. It can be written as $\mathcal{H}\equiv\mathcal{H}_{comp}\oplus \mathcal{H}_{err}$ with $\mathcal{H}_{comp}\equiv \{0,1\}^{\otimes n} \otimes \{\ket{\Psi_{s}}\}$.  A logical operation $L_j$ at time step $j$ in the ECC consists of a subsequence $G_j$ of (faulty) local gates, followed by syndrome measurement $M_j$ in $\mathcal{H}_{err}$ and correction feedback $F_j$ towards $\mathcal{H}_{comp}$, or $L_j=F_jM_jG_j$. As before, to do randomized benchmarking of one component we require that no phase relationship is built up with the second, for which we use a generalized operation $R_j$. The one-design sequence can then be written as 
\begin{align}
\Gamma_y &=\frac{1}{\sharp\mathcal{C}^y}\sum_{\{C_j\}\in\mathcal{C}^y}\PTrace{\rm comp}{ \left(\prod_{j=y}^1(\Lambda_j R_j C_j)\right)\hat{\rho}_0}
\end{align}
where $C_j$ are logically-encoded Clifford gates prior to error correction.  As with the general protocol and the examples discussed above, we again ensure we take the correct group average that reduces our error channel to a depolarizing one; in the present case we do so using the half-twirl using the Clifford group amended with phase randomization between the  $\mathcal{H}_{err}$ and $\mathcal{H}_{comp}$ subspaces.  There are actually different ways to do this. As with the number conservation, we can do it with single-qubit phase gates on the physical qubits, i.e.~$R_j=Z^{\otimes m}_j$. Another way to randomize the phase is with syndrome measurements, i.e.~$R_j=M_j$.  Yet another (more subtle) way is to apply both a syndrome measurement (throwing away the result) and a randomly chosen correction operation $F_j$ from amongst the possible correction operations $\{F_j\}$, so that  $R_j=F_jM_j$.  At the end, one obtains an average error per gate estimate of the compound operation $RC$,~i.e.~$\mu_{RC}$. Since we can obtain the single-qubit error rates ($\mu_Z$) from standard randomized benchmarking of the logical Clifford operations \cite{Combes2017}, we can simply derive estimates for the other components as $\mu_{C}\approx\mu_{RC}-\mu_{R}$, as with standard interleaved benchmarking. 

Using the above protocol, we now have a means to obtain the decay rate into the  $\mathcal{H}_{err}$ subspace.  Note that this does not correspond to a logical error since the vast majority of these events are suppressed by the error correction feedback.  Nonetheless, we can use the protocol with interleaved gates to benchmark arbitrary logical operations $G$ with respect to this error channel, obtaining $\mu_G$.  If we further make the standard assumption from ECC that correlated errors longer than the distance $d$ of the code are negligible, we can then simply upperbound the logical error rate corresponding to computational faults as $\mu^L_G<(\mu_{F}+\mu_{M}+\mu_{G})^d$.
Note also that the above considerations could be particularly insightful for concatenated error codes, where the logical error rate at one layer corresponds to the physical error rate at the layer above. For concatenated codes as well as for other ECC protocols, benchmarking non-Clifford gates is pivotal to fault-tolerant universal quantum computation.

\section{Conclusion}
Both quantum simulation and universal quantum computation involve complex processes that cannot be efficiently predicted classically.  We show that this restriction does not prohibit their validation, provided the implementation being benchmarked can be found or engineered to conserve symmetries in the system. Such is the case for many quantum simulation tasks, such as fermionic systems and the Fermi-Hubbard model, as well as for fault-tolerant quantum computation, where stabilizers of error correcting codes are preserved by logical operations. We present a symmetry benchmarking protocol relying on randomization via unitary one--designs on conserved subspaces, that allows extraction of average channel error while maintaining robustness to state preparation and measurement imperfections

\section{Acknowledgements}
We acknowledge funding through SCALEQIT and a Google Faculty Research Award.

\bibliography{LRB}

\begin{thebibliography}{40}%
\makeatletter
\providecommand \@ifxundefined [1]{%
 \@ifx{#1\undefined}
}%
\providecommand \@ifnum [1]{%
 \ifnum #1\expandafter \@firstoftwo
 \else \expandafter \@secondoftwo
 \fi
}%
\providecommand \@ifx [1]{%
 \ifx #1\expandafter \@firstoftwo
 \else \expandafter \@secondoftwo
 \fi
}%
\providecommand \natexlab [1]{#1}%
\providecommand \enquote  [1]{``#1''}%
\providecommand \bibnamefont  [1]{#1}%
\providecommand \bibfnamefont [1]{#1}%
\providecommand \citenamefont [1]{#1}%
\providecommand \href@noop [0]{\@secondoftwo}%
\providecommand \href [0]{\begingroup \@sanitize@url \@href}%
\providecommand \@href[1]{\@@startlink{#1}\@@href}%
\providecommand \@@href[1]{\endgroup#1\@@endlink}%
\providecommand \@sanitize@url [0]{\catcode `\\12\catcode `\$12\catcode
  `\&12\catcode `\#12\catcode `\^12\catcode `\_12\catcode `\%12\relax}%
\providecommand \@@startlink[1]{}%
\providecommand \@@endlink[0]{}%
\providecommand \url  [0]{\begingroup\@sanitize@url \@url }%
\providecommand \@url [1]{\endgroup\@href {#1}{\urlprefix }}%
\providecommand \urlprefix  [0]{URL }%
\providecommand \Eprint [0]{\href }%
\providecommand \doibase [0]{http://dx.doi.org/}%
\providecommand \selectlanguage [0]{\@gobble}%
\providecommand \bibinfo  [0]{\@secondoftwo}%
\providecommand \bibfield  [0]{\@secondoftwo}%
\providecommand \translation [1]{[#1]}%
\providecommand \BibitemOpen [0]{}%
\providecommand \bibitemStop [0]{}%
\providecommand \bibitemNoStop [0]{.\EOS\space}%
\providecommand \EOS [0]{\spacefactor3000\relax}%
\providecommand \BibitemShut  [1]{\csname bibitem#1\endcsname}%
\let\auto@bib@innerbib\@empty
%</preamble>
\bibitem [{\citenamefont {Bell}(1964)}]{bell1964einstein}%
  \BibitemOpen
  \bibfield  {author} {\bibinfo {author} {\bibfnamefont {J.~S.}\ \bibnamefont
  {Bell}},\ }\href@noop {} {\  (\bibinfo {year} {1964})}\BibitemShut {NoStop}%
\bibitem [{\citenamefont {Aspect}\ \emph {et~al.}(1982)\citenamefont {Aspect},
  \citenamefont {Dalibard},\ and\ \citenamefont
  {Roger}}]{aspect1982experimental}%
  \BibitemOpen
  \bibfield  {author} {\bibinfo {author} {\bibfnamefont {A.}~\bibnamefont
  {Aspect}}, \bibinfo {author} {\bibfnamefont {J.}~\bibnamefont {Dalibard}}, \
  and\ \bibinfo {author} {\bibfnamefont {G.}~\bibnamefont {Roger}},\
  }\href@noop {} {\bibfield  {journal} {\bibinfo  {journal} {Physical review
  letters}\ }\textbf {\bibinfo {volume} {49}},\ \bibinfo {pages} {1804}
  (\bibinfo {year} {1982})}\BibitemShut {NoStop}%
\bibitem [{\citenamefont {Goldreich}(2010)}]{goldreich2010p}%
  \BibitemOpen
  \bibfield  {author} {\bibinfo {author} {\bibfnamefont {O.}~\bibnamefont
  {Goldreich}},\ }\href@noop {} {\emph {\bibinfo {title} {P, NP, and
  NP-Completeness: The basics of computational complexity}}}\ (\bibinfo
  {publisher} {Cambridge University Press},\ \bibinfo {year}
  {2010})\BibitemShut {NoStop}%
\bibitem [{\citenamefont {Gross}\ \emph {et~al.}(2010)\citenamefont {Gross},
  \citenamefont {Liu}, \citenamefont {Flammia}, \citenamefont {Becker},\ and\
  \citenamefont {Eisert}}]{gross2010quantum}%
  \BibitemOpen
  \bibfield  {author} {\bibinfo {author} {\bibfnamefont {D.}~\bibnamefont
  {Gross}}, \bibinfo {author} {\bibfnamefont {Y.-K.}\ \bibnamefont {Liu}},
  \bibinfo {author} {\bibfnamefont {S.~T.}\ \bibnamefont {Flammia}}, \bibinfo
  {author} {\bibfnamefont {S.}~\bibnamefont {Becker}}, \ and\ \bibinfo {author}
  {\bibfnamefont {J.}~\bibnamefont {Eisert}},\ }\href@noop {} {\bibfield
  {journal} {\bibinfo  {journal} {Physical review letters}\ }\textbf {\bibinfo
  {volume} {105}},\ \bibinfo {pages} {150401} (\bibinfo {year}
  {2010})}\BibitemShut {NoStop}%
\bibitem [{\citenamefont {Cramer}\ \emph {et~al.}(2010)\citenamefont {Cramer},
  \citenamefont {Plenio}, \citenamefont {Flammia}, \citenamefont {Somma},
  \citenamefont {Gross}, \citenamefont {Bartlett}, \citenamefont
  {Landon-Cardinal}, \citenamefont {Poulin},\ and\ \citenamefont
  {Liu}}]{cramer2010efficient}%
  \BibitemOpen
  \bibfield  {author} {\bibinfo {author} {\bibfnamefont {M.}~\bibnamefont
  {Cramer}}, \bibinfo {author} {\bibfnamefont {M.~B.}\ \bibnamefont {Plenio}},
  \bibinfo {author} {\bibfnamefont {S.~T.}\ \bibnamefont {Flammia}}, \bibinfo
  {author} {\bibfnamefont {R.}~\bibnamefont {Somma}}, \bibinfo {author}
  {\bibfnamefont {D.}~\bibnamefont {Gross}}, \bibinfo {author} {\bibfnamefont
  {S.~D.}\ \bibnamefont {Bartlett}}, \bibinfo {author} {\bibfnamefont
  {O.}~\bibnamefont {Landon-Cardinal}}, \bibinfo {author} {\bibfnamefont
  {D.}~\bibnamefont {Poulin}}, \ and\ \bibinfo {author} {\bibfnamefont {Y.-K.}\
  \bibnamefont {Liu}},\ }\href@noop {} {\bibfield  {journal} {\bibinfo
  {journal} {Nature communications}\ }\textbf {\bibinfo {volume} {1}},\
  \bibinfo {pages} {149} (\bibinfo {year} {2010})}\BibitemShut {NoStop}%
\bibitem [{\citenamefont {Gross}(2011)}]{gross2011recovering}%
  \BibitemOpen
  \bibfield  {author} {\bibinfo {author} {\bibfnamefont {D.}~\bibnamefont
  {Gross}},\ }\href@noop {} {\bibfield  {journal} {\bibinfo  {journal} {IEEE
  Transactions on Information Theory}\ }\textbf {\bibinfo {volume} {57}},\
  \bibinfo {pages} {1548} (\bibinfo {year} {2011})}\BibitemShut {NoStop}%
\bibitem [{\citenamefont {Shabani}\ \emph {et~al.}(2011)\citenamefont
  {Shabani}, \citenamefont {Kosut}, \citenamefont {Mohseni}, \citenamefont
  {Rabitz}, \citenamefont {Broome}, \citenamefont {Almeida}, \citenamefont
  {Fedrizzi},\ and\ \citenamefont {White}}]{shabani2011efficient}%
  \BibitemOpen
  \bibfield  {author} {\bibinfo {author} {\bibfnamefont {A.}~\bibnamefont
  {Shabani}}, \bibinfo {author} {\bibfnamefont {R.}~\bibnamefont {Kosut}},
  \bibinfo {author} {\bibfnamefont {M.}~\bibnamefont {Mohseni}}, \bibinfo
  {author} {\bibfnamefont {H.}~\bibnamefont {Rabitz}}, \bibinfo {author}
  {\bibfnamefont {M.}~\bibnamefont {Broome}}, \bibinfo {author} {\bibfnamefont
  {M.}~\bibnamefont {Almeida}}, \bibinfo {author} {\bibfnamefont
  {A.}~\bibnamefont {Fedrizzi}}, \ and\ \bibinfo {author} {\bibfnamefont
  {A.}~\bibnamefont {White}},\ }\href@noop {} {\bibfield  {journal} {\bibinfo
  {journal} {Physical review letters}\ }\textbf {\bibinfo {volume} {106}},\
  \bibinfo {pages} {100401} (\bibinfo {year} {2011})}\BibitemShut {NoStop}%
\bibitem [{\citenamefont {Knill}\ \emph {et~al.}(2008)\citenamefont {Knill},
  \citenamefont {Leibfried}, \citenamefont {Reichle}, \citenamefont {Britton},
  \citenamefont {Blakestad}, \citenamefont {Jost}, \citenamefont {Langer},
  \citenamefont {Ozeri}, \citenamefont {Seidelin},\ and\ \citenamefont
  {Wineland}}]{Knill2008}%
  \BibitemOpen
  \bibfield  {author} {\bibinfo {author} {\bibfnamefont {E.}~\bibnamefont
  {Knill}}, \bibinfo {author} {\bibfnamefont {D.}~\bibnamefont {Leibfried}},
  \bibinfo {author} {\bibfnamefont {R.}~\bibnamefont {Reichle}}, \bibinfo
  {author} {\bibfnamefont {J.}~\bibnamefont {Britton}}, \bibinfo {author}
  {\bibfnamefont {R.}~\bibnamefont {Blakestad}}, \bibinfo {author}
  {\bibfnamefont {J.}~\bibnamefont {Jost}}, \bibinfo {author} {\bibfnamefont
  {C.}~\bibnamefont {Langer}}, \bibinfo {author} {\bibfnamefont
  {R.}~\bibnamefont {Ozeri}}, \bibinfo {author} {\bibfnamefont
  {S.}~\bibnamefont {Seidelin}}, \ and\ \bibinfo {author} {\bibfnamefont
  {D.}~\bibnamefont {Wineland}},\ }\href@noop {} {\bibfield  {journal}
  {\bibinfo  {journal} {Phys. Rev. A}\ }\textbf {\bibinfo {volume} {77}},\
  \bibinfo {pages} {012307} (\bibinfo {year} {2008})}\BibitemShut {NoStop}%
\bibitem [{\citenamefont {Magesan}\ \emph {et~al.}(2011)\citenamefont
  {Magesan}, \citenamefont {Gambetta},\ and\ \citenamefont
  {Emerson}}]{Magesan2011}%
  \BibitemOpen
  \bibfield  {author} {\bibinfo {author} {\bibfnamefont {E.}~\bibnamefont
  {Magesan}}, \bibinfo {author} {\bibfnamefont {J.}~\bibnamefont {Gambetta}}, \
  and\ \bibinfo {author} {\bibfnamefont {J.}~\bibnamefont {Emerson}},\
  }\href@noop {} {\bibfield  {journal} {\bibinfo  {journal} {Phys. Rev. Lett.}\
  }\textbf {\bibinfo {volume} {106}},\ \bibinfo {pages} {180504} (\bibinfo
  {year} {2011})}\BibitemShut {NoStop}%
\bibitem [{\citenamefont {Nielsen}\ and\ \citenamefont
  {Chuang}(2000)}]{Nielsen2000}%
  \BibitemOpen
  \bibfield  {author} {\bibinfo {author} {\bibfnamefont {M.}~\bibnamefont
  {Nielsen}}\ and\ \bibinfo {author} {\bibfnamefont {I.}~\bibnamefont
  {Chuang}},\ }\href@noop {} {\emph {\bibinfo {title} {Quantum Computation and
  Quantum Information}}}\ (\bibinfo  {publisher} {Cambridge University Press},\
  \bibinfo {year} {2000})\BibitemShut {NoStop}%
\bibitem [{\citenamefont {Farhi}\ \emph {et~al.}(2001)\citenamefont {Farhi},
  \citenamefont {Goldstone}, \citenamefont {Gutmann}, \citenamefont {Lapan},
  \citenamefont {Lundgren},\ and\ \citenamefont {Preda}}]{farhi2001quantum}%
  \BibitemOpen
  \bibfield  {author} {\bibinfo {author} {\bibfnamefont {E.}~\bibnamefont
  {Farhi}}, \bibinfo {author} {\bibfnamefont {J.}~\bibnamefont {Goldstone}},
  \bibinfo {author} {\bibfnamefont {S.}~\bibnamefont {Gutmann}}, \bibinfo
  {author} {\bibfnamefont {J.}~\bibnamefont {Lapan}}, \bibinfo {author}
  {\bibfnamefont {A.}~\bibnamefont {Lundgren}}, \ and\ \bibinfo {author}
  {\bibfnamefont {D.}~\bibnamefont {Preda}},\ }\href@noop {} {\bibfield
  {journal} {\bibinfo  {journal} {Science}\ }\textbf {\bibinfo {volume}
  {292}},\ \bibinfo {pages} {472} (\bibinfo {year} {2001})}\BibitemShut
  {NoStop}%
\bibitem [{\citenamefont {Lloyd}(1996)}]{lloyd1996universal}%
  \BibitemOpen
  \bibfield  {author} {\bibinfo {author} {\bibfnamefont {S.}~\bibnamefont
  {Lloyd}},\ }\href@noop {} {\bibfield  {journal} {\bibinfo  {journal}
  {Science}\ }\textbf {\bibinfo {volume} {273}},\ \bibinfo {pages} {1073}
  (\bibinfo {year} {1996})}\BibitemShut {NoStop}%
\bibitem [{\citenamefont {Aspuru-Guzik}\ \emph {et~al.}(2005)\citenamefont
  {Aspuru-Guzik}, \citenamefont {Dutoi}, \citenamefont {Love},\ and\
  \citenamefont {Head-Gordon}}]{aspuru2005simulated}%
  \BibitemOpen
  \bibfield  {author} {\bibinfo {author} {\bibfnamefont {A.}~\bibnamefont
  {Aspuru-Guzik}}, \bibinfo {author} {\bibfnamefont {A.~D.}\ \bibnamefont
  {Dutoi}}, \bibinfo {author} {\bibfnamefont {P.~J.}\ \bibnamefont {Love}}, \
  and\ \bibinfo {author} {\bibfnamefont {M.}~\bibnamefont {Head-Gordon}},\
  }\href@noop {} {\bibfield  {journal} {\bibinfo  {journal} {Science}\ }\textbf
  {\bibinfo {volume} {309}},\ \bibinfo {pages} {1704} (\bibinfo {year}
  {2005})}\BibitemShut {NoStop}%
\bibitem [{\citenamefont {Kassal}\ \emph {et~al.}(2011)\citenamefont {Kassal},
  \citenamefont {Whitfield}, \citenamefont {Perdomo-Ortiz}, \citenamefont
  {Yung},\ and\ \citenamefont {Aspuru-Guzik}}]{Kassal2011}%
  \BibitemOpen
  \bibfield  {author} {\bibinfo {author} {\bibfnamefont {I.}~\bibnamefont
  {Kassal}}, \bibinfo {author} {\bibfnamefont {J.~D.}\ \bibnamefont
  {Whitfield}}, \bibinfo {author} {\bibfnamefont {A.}~\bibnamefont
  {Perdomo-Ortiz}}, \bibinfo {author} {\bibfnamefont {M.-H.}\ \bibnamefont
  {Yung}}, \ and\ \bibinfo {author} {\bibfnamefont {A.}~\bibnamefont
  {Aspuru-Guzik}},\ }\href {\doibase
  http://dx.doi.org/10.1146/annurev-physchem-032210-103512} {\bibfield
  {journal} {\bibinfo  {journal} {Annual Review of Physical Chemistry}\
  }\textbf {\bibinfo {volume} {62}},\ \bibinfo {pages} {185} (\bibinfo {year}
  {2011})}\BibitemShut {NoStop}%
\bibitem [{\citenamefont {McClean}\ \emph {et~al.}(2016)\citenamefont
  {McClean}, \citenamefont {Romero}, \citenamefont {Babbush},\ and\
  \citenamefont {Aspuru-Guzik}}]{mcclean2016theory}%
  \BibitemOpen
  \bibfield  {author} {\bibinfo {author} {\bibfnamefont {J.~R.}\ \bibnamefont
  {McClean}}, \bibinfo {author} {\bibfnamefont {J.}~\bibnamefont {Romero}},
  \bibinfo {author} {\bibfnamefont {R.}~\bibnamefont {Babbush}}, \ and\
  \bibinfo {author} {\bibfnamefont {A.}~\bibnamefont {Aspuru-Guzik}},\
  }\href@noop {} {\bibfield  {journal} {\bibinfo  {journal} {New Journal of
  Physics}\ }\textbf {\bibinfo {volume} {18}},\ \bibinfo {pages} {023023}
  (\bibinfo {year} {2016})}\BibitemShut {NoStop}%
\bibitem [{\citenamefont {Barends}\ \emph {et~al.}(2015)\citenamefont
  {Barends}, \citenamefont {Lamata}, \citenamefont {Kelly}, \citenamefont
  {Garcia-Alvarez}, \citenamefont {Fowler}, \citenamefont {Megrant},
  \citenamefont {Jeffrey}, \citenamefont {White}, \citenamefont {Sank},
  \citenamefont {Mutus}, \citenamefont {Campbell}, \citenamefont {Chen},
  \citenamefont {Chen}, \citenamefont {Chiaro}, \citenamefont {Dunsworth},
  \citenamefont {Hoi}, \citenamefont {Neill}, \citenamefont {O'Malley},
  \citenamefont {Quintana}, \citenamefont {Roushan}, \citenamefont
  {Vainsencher}, \citenamefont {Wenner}, \citenamefont {Solano},\ and\
  \citenamefont {Martinis}}]{Barends2015}%
  \BibitemOpen
  \bibfield  {author} {\bibinfo {author} {\bibfnamefont {R.}~\bibnamefont
  {Barends}}, \bibinfo {author} {\bibfnamefont {L.}~\bibnamefont {Lamata}},
  \bibinfo {author} {\bibfnamefont {J.}~\bibnamefont {Kelly}}, \bibinfo
  {author} {\bibfnamefont {L.}~\bibnamefont {Garcia-Alvarez}}, \bibinfo
  {author} {\bibfnamefont {A.~G.}\ \bibnamefont {Fowler}}, \bibinfo {author}
  {\bibfnamefont {A.}~\bibnamefont {Megrant}}, \bibinfo {author} {\bibfnamefont
  {E.}~\bibnamefont {Jeffrey}}, \bibinfo {author} {\bibfnamefont {T.~C.}\
  \bibnamefont {White}}, \bibinfo {author} {\bibfnamefont {D.}~\bibnamefont
  {Sank}}, \bibinfo {author} {\bibfnamefont {J.~Y.}\ \bibnamefont {Mutus}},
  \bibinfo {author} {\bibfnamefont {B.}~\bibnamefont {Campbell}}, \bibinfo
  {author} {\bibfnamefont {Y.}~\bibnamefont {Chen}}, \bibinfo {author}
  {\bibfnamefont {Z.}~\bibnamefont {Chen}}, \bibinfo {author} {\bibfnamefont
  {B.}~\bibnamefont {Chiaro}}, \bibinfo {author} {\bibfnamefont
  {A.}~\bibnamefont {Dunsworth}}, \bibinfo {author} {\bibfnamefont {I.-C.}\
  \bibnamefont {Hoi}}, \bibinfo {author} {\bibfnamefont {C.}~\bibnamefont
  {Neill}}, \bibinfo {author} {\bibfnamefont {P.~J.~J.}\ \bibnamefont
  {O'Malley}}, \bibinfo {author} {\bibfnamefont {C.}~\bibnamefont {Quintana}},
  \bibinfo {author} {\bibfnamefont {P.}~\bibnamefont {Roushan}}, \bibinfo
  {author} {\bibfnamefont {A.}~\bibnamefont {Vainsencher}}, \bibinfo {author}
  {\bibfnamefont {J.}~\bibnamefont {Wenner}}, \bibinfo {author} {\bibfnamefont
  {E.}~\bibnamefont {Solano}}, \ and\ \bibinfo {author} {\bibfnamefont {J.~M.}\
  \bibnamefont {Martinis}},\ }\href {\doibase
  http://dx.doi.org/10.1038/ncomms8654} {\bibfield  {journal} {\bibinfo
  {journal} {Nature Communications}\ }\textbf {\bibinfo {volume} {6}},\
  \bibinfo {pages} {7654} (\bibinfo {year} {2015})}\BibitemShut {NoStop}%
\bibitem [{\citenamefont {Wecker}\ \emph {et~al.}(2015)\citenamefont {Wecker},
  \citenamefont {Hastings}, \citenamefont {Wiebe}, \citenamefont {Clark},
  \citenamefont {Nayak},\ and\ \citenamefont {Troyer}}]{Wecker2015a}%
  \BibitemOpen
  \bibfield  {author} {\bibinfo {author} {\bibfnamefont {D.}~\bibnamefont
  {Wecker}}, \bibinfo {author} {\bibfnamefont {M.~B.}\ \bibnamefont
  {Hastings}}, \bibinfo {author} {\bibfnamefont {N.}~\bibnamefont {Wiebe}},
  \bibinfo {author} {\bibfnamefont {B.~K.}\ \bibnamefont {Clark}}, \bibinfo
  {author} {\bibfnamefont {C.}~\bibnamefont {Nayak}}, \ and\ \bibinfo {author}
  {\bibfnamefont {M.}~\bibnamefont {Troyer}},\ }\href {\doibase
  http://dx.doi.org/10.1103/PhysRevA.92.062318} {\bibfield  {journal} {\bibinfo
   {journal} {Phys. Rev. A}\ }\textbf {\bibinfo {volume} {92}},\ \bibinfo
  {pages} {062318} (\bibinfo {year} {2015})}\BibitemShut {NoStop}%
\bibitem [{\citenamefont {Bauer}\ \emph {et~al.}(2016)\citenamefont {Bauer},
  \citenamefont {Wecker}, \citenamefont {Millis}, \citenamefont {Hastings},\
  and\ \citenamefont {Troyer}}]{Bauer2016}%
  \BibitemOpen
  \bibfield  {author} {\bibinfo {author} {\bibfnamefont {B.}~\bibnamefont
  {Bauer}}, \bibinfo {author} {\bibfnamefont {D.}~\bibnamefont {Wecker}},
  \bibinfo {author} {\bibfnamefont {A.~J.}\ \bibnamefont {Millis}}, \bibinfo
  {author} {\bibfnamefont {M.~B.}\ \bibnamefont {Hastings}}, \ and\ \bibinfo
  {author} {\bibfnamefont {M.}~\bibnamefont {Troyer}},\ }\href {\doibase
  10.1103/physrevx.6.031045} {\bibfield  {journal} {\bibinfo  {journal}
  {Physical Review X}\ }\textbf {\bibinfo {volume} {6}} (\bibinfo {year}
  {2016}),\ 10.1103/physrevx.6.031045}\BibitemShut {NoStop}%
\bibitem [{\citenamefont {Dallaire-Demers}\ and\ \citenamefont
  {Wilhelm}(2016{\natexlab{a}})}]{DallaireDemers2016b}%
  \BibitemOpen
  \bibfield  {author} {\bibinfo {author} {\bibfnamefont {P.-L.}\ \bibnamefont
  {Dallaire-Demers}}\ and\ \bibinfo {author} {\bibfnamefont {F.~K.}\
  \bibnamefont {Wilhelm}},\ }\href {\doibase
  https://doi.org/10.1103/PhysRevA.94.062304} {\bibfield  {journal} {\bibinfo
  {journal} {Phys. Rev. A}\ }\textbf {\bibinfo {volume} {94}},\ \bibinfo
  {pages} {062304} (\bibinfo {year} {2016}{\natexlab{a}})}\BibitemShut
  {NoStop}%
\bibitem [{\citenamefont {Salathe}\ \emph {et~al.}(2015)\citenamefont
  {Salathe}, \citenamefont {Mondal}, \citenamefont {Oppliger}, \citenamefont
  {Heinsoo}, \citenamefont {Kurpiers}, \citenamefont {Potocnik}, \citenamefont
  {Mezzacapo}, \citenamefont {Heras}, \citenamefont {Lamata}, \citenamefont
  {Solano}, \citenamefont {Filipp},\ and\ \citenamefont
  {Wallraff}}]{Salathe2015}%
  \BibitemOpen
  \bibfield  {author} {\bibinfo {author} {\bibfnamefont {Y.}~\bibnamefont
  {Salathe}}, \bibinfo {author} {\bibfnamefont {M.}~\bibnamefont {Mondal}},
  \bibinfo {author} {\bibfnamefont {M.}~\bibnamefont {Oppliger}}, \bibinfo
  {author} {\bibfnamefont {J.}~\bibnamefont {Heinsoo}}, \bibinfo {author}
  {\bibfnamefont {P.}~\bibnamefont {Kurpiers}}, \bibinfo {author}
  {\bibfnamefont {A.}~\bibnamefont {Potocnik}}, \bibinfo {author}
  {\bibfnamefont {A.}~\bibnamefont {Mezzacapo}}, \bibinfo {author}
  {\bibfnamefont {U.~L.}\ \bibnamefont {Heras}}, \bibinfo {author}
  {\bibfnamefont {L.}~\bibnamefont {Lamata}}, \bibinfo {author} {\bibfnamefont
  {E.}~\bibnamefont {Solano}}, \bibinfo {author} {\bibfnamefont
  {S.}~\bibnamefont {Filipp}}, \ and\ \bibinfo {author} {\bibfnamefont
  {A.}~\bibnamefont {Wallraff}},\ }\href {\doibase
  http://dx.doi.org/10.1103/PhysRevX.5.021027} {\bibfield  {journal} {\bibinfo
  {journal} {Phys. Rev. X}\ }\textbf {\bibinfo {volume} {5}},\ \bibinfo {pages}
  {021027} (\bibinfo {year} {2015})}\BibitemShut {NoStop}%
\bibitem [{\citenamefont {Martinez}\ \emph {et~al.}(2016)\citenamefont
  {Martinez}, \citenamefont {Muschik}, \citenamefont {Schindler}, \citenamefont
  {Nigg}, \citenamefont {Erhard}, \citenamefont {Heyl}, \citenamefont {Hauke},
  \citenamefont {Dalmonte}, \citenamefont {Monz}, \citenamefont {Zoller},\ and\
  \citenamefont {Blatt}}]{Martinez2016}%
  \BibitemOpen
  \bibfield  {author} {\bibinfo {author} {\bibfnamefont {E.~A.}\ \bibnamefont
  {Martinez}}, \bibinfo {author} {\bibfnamefont {C.~A.}\ \bibnamefont
  {Muschik}}, \bibinfo {author} {\bibfnamefont {P.}~\bibnamefont {Schindler}},
  \bibinfo {author} {\bibfnamefont {D.}~\bibnamefont {Nigg}}, \bibinfo {author}
  {\bibfnamefont {A.}~\bibnamefont {Erhard}}, \bibinfo {author} {\bibfnamefont
  {M.}~\bibnamefont {Heyl}}, \bibinfo {author} {\bibfnamefont {P.}~\bibnamefont
  {Hauke}}, \bibinfo {author} {\bibfnamefont {M.}~\bibnamefont {Dalmonte}},
  \bibinfo {author} {\bibfnamefont {T.}~\bibnamefont {Monz}}, \bibinfo {author}
  {\bibfnamefont {P.}~\bibnamefont {Zoller}}, \ and\ \bibinfo {author}
  {\bibfnamefont {R.}~\bibnamefont {Blatt}},\ }\href {\doibase
  http://dx.doi.org/10.1038/nature18318} {\bibfield  {journal} {\bibinfo
  {journal} {Nature}\ }\textbf {\bibinfo {volume} {534}},\ \bibinfo {pages}
  {516} (\bibinfo {year} {2016})}\BibitemShut {NoStop}%
\bibitem [{\citenamefont {Zohar}\ \emph {et~al.}(2017)\citenamefont {Zohar},
  \citenamefont {Farace}, \citenamefont {Reznik},\ and\ \citenamefont
  {Cirac}}]{Zohar2017}%
  \BibitemOpen
  \bibfield  {author} {\bibinfo {author} {\bibfnamefont {E.}~\bibnamefont
  {Zohar}}, \bibinfo {author} {\bibfnamefont {A.}~\bibnamefont {Farace}},
  \bibinfo {author} {\bibfnamefont {B.}~\bibnamefont {Reznik}}, \ and\ \bibinfo
  {author} {\bibfnamefont {J.~I.}\ \bibnamefont {Cirac}},\ }\href {\doibase
  http://dx.doi.org/10.1103/PhysRevA.95.023604} {\bibfield  {journal} {\bibinfo
   {journal} {Physical Review A}\ }\textbf {\bibinfo {volume} {95}} (\bibinfo
  {year} {2017}),\ http://dx.doi.org/10.1103/PhysRevA.95.023604}\BibitemShut
  {NoStop}%
\bibitem [{\citenamefont {Magesan}\ \emph {et~al.}(2012)\citenamefont
  {Magesan}, \citenamefont {Gambetta}, \citenamefont {Johnson}, \citenamefont
  {Ryan}, \citenamefont {Chow}, \citenamefont {Merkel}, \citenamefont
  {da~Silva}, \citenamefont {Keefe}, \citenamefont {Rothwell}, \citenamefont
  {Ohki}, \citenamefont {Ketchen},\ and\ \citenamefont
  {Steffen}}]{Magesan2012}%
  \BibitemOpen
  \bibfield  {author} {\bibinfo {author} {\bibfnamefont {E.}~\bibnamefont
  {Magesan}}, \bibinfo {author} {\bibfnamefont {J.}~\bibnamefont {Gambetta}},
  \bibinfo {author} {\bibfnamefont {B.~R.}\ \bibnamefont {Johnson}}, \bibinfo
  {author} {\bibfnamefont {C.}~\bibnamefont {Ryan}}, \bibinfo {author}
  {\bibfnamefont {J.}~\bibnamefont {Chow}}, \bibinfo {author} {\bibfnamefont
  {S.}~\bibnamefont {Merkel}}, \bibinfo {author} {\bibfnamefont
  {M.}~\bibnamefont {da~Silva}}, \bibinfo {author} {\bibfnamefont
  {G.}~\bibnamefont {Keefe}}, \bibinfo {author} {\bibfnamefont
  {M.}~\bibnamefont {Rothwell}}, \bibinfo {author} {\bibfnamefont
  {T.}~\bibnamefont {Ohki}}, \bibinfo {author} {\bibfnamefont {M.}~\bibnamefont
  {Ketchen}}, \ and\ \bibinfo {author} {\bibfnamefont {M.}~\bibnamefont
  {Steffen}},\ }\href@noop {} {\bibfield  {journal} {\bibinfo  {journal} {Phys.
  Rev. Lett.}\ }\textbf {\bibinfo {volume} {109}},\ \bibinfo {pages} {080505}
  (\bibinfo {year} {2012})}\BibitemShut {NoStop}%
\bibitem [{\citenamefont {Chasseur}\ and\ \citenamefont
  {Wilhelm}(2015)}]{Chasseur2015}%
  \BibitemOpen
  \bibfield  {author} {\bibinfo {author} {\bibfnamefont {T.}~\bibnamefont
  {Chasseur}}\ and\ \bibinfo {author} {\bibfnamefont {F.}~\bibnamefont
  {Wilhelm}},\ }\href@noop {} {\bibfield  {journal} {\bibinfo  {journal} {Phys.
  Rev. A}\ }\textbf {\bibinfo {volume} {92}},\ \bibinfo {pages} {042333}
  (\bibinfo {year} {2015})}\BibitemShut {NoStop}%
\bibitem [{\citenamefont {Wallman}\ \emph {et~al.}(2014)\citenamefont
  {Wallman}, \citenamefont {Barnhill},\ and\ \citenamefont
  {Emerson}}]{Wallman2014X}%
  \BibitemOpen
  \bibfield  {author} {\bibinfo {author} {\bibfnamefont {J.}~\bibnamefont
  {Wallman}}, \bibinfo {author} {\bibfnamefont {M.}~\bibnamefont {Barnhill}}, \
  and\ \bibinfo {author} {\bibfnamefont {J.}~\bibnamefont {Emerson}},\
  }\href@noop {} {\enquote {\bibinfo {title} {Characterization of leakage
  errors via randomized benchmarking},}\ } (\bibinfo {year} {2014}),\ \bibinfo
  {note} {arXiv:1412.4126}\BibitemShut {NoStop}%
\bibitem [{\citenamefont {Carignan-Dugas}\ \emph {et~al.}(2015)\citenamefont
  {Carignan-Dugas}, \citenamefont {Wallman},\ and\ \citenamefont
  {Emerson}}]{Carignan-Dougas2015}%
  \BibitemOpen
  \bibfield  {author} {\bibinfo {author} {\bibfnamefont {A.}~\bibnamefont
  {Carignan-Dugas}}, \bibinfo {author} {\bibfnamefont {J.}~\bibnamefont
  {Wallman}}, \ and\ \bibinfo {author} {\bibfnamefont {J.}~\bibnamefont
  {Emerson}},\ }\href@noop {} {\bibfield  {journal} {\bibinfo  {journal} {Phys.
  Rev. A}\ }\textbf {\bibinfo {volume} {92}},\ \bibinfo {pages} {060302(R)}
  (\bibinfo {year} {2015})}\BibitemShut {NoStop}%
\bibitem [{\citenamefont {Cross}\ \emph {et~al.}(2016)\citenamefont {Cross},
  \citenamefont {Magesan}, \citenamefont {Bishop}, \citenamefont {Smolin},\
  and\ \citenamefont {Gambetta}}]{Cross2016}%
  \BibitemOpen
  \bibfield  {author} {\bibinfo {author} {\bibfnamefont {A.}~\bibnamefont
  {Cross}}, \bibinfo {author} {\bibfnamefont {E.}~\bibnamefont {Magesan}},
  \bibinfo {author} {\bibfnamefont {L.}~\bibnamefont {Bishop}}, \bibinfo
  {author} {\bibfnamefont {J.}~\bibnamefont {Smolin}}, \ and\ \bibinfo {author}
  {\bibfnamefont {J.}~\bibnamefont {Gambetta}},\ }\href@noop {} {\bibfield
  {journal} {\bibinfo  {journal} {npjqi}\ }\textbf {\bibinfo {volume} {2}},\
  \bibinfo {pages} {16012} (\bibinfo {year} {2016})}\BibitemShut {NoStop}%
\bibitem [{\citenamefont {Chasseur}\ \emph {et~al.}(2016)\citenamefont
  {Chasseur}, \citenamefont {Reich}, \citenamefont {Wilhelm},\ and\
  \citenamefont {Koch}}]{Chasseur2016}%
  \BibitemOpen
  \bibfield  {author} {\bibinfo {author} {\bibfnamefont {T.}~\bibnamefont
  {Chasseur}}, \bibinfo {author} {\bibfnamefont {D.}~\bibnamefont {Reich}},
  \bibinfo {author} {\bibfnamefont {F.}~\bibnamefont {Wilhelm}}, \ and\
  \bibinfo {author} {\bibfnamefont {C.}~\bibnamefont {Koch}},\ }\href@noop {}
  {\enquote {\bibinfo {title} {Hybrid benchmarking of arbitrary quantum
  gates},}\ } (\bibinfo {year} {2016}),\ \bibinfo {note}
  {arXiv:1606.03927}\BibitemShut {NoStop}%
\bibitem [{\citenamefont {Reich}\ \emph {et~al.}(2013)\citenamefont {Reich},
  \citenamefont {Gualdi},\ and\ \citenamefont {Koch}}]{Reich2013}%
  \BibitemOpen
  \bibfield  {author} {\bibinfo {author} {\bibfnamefont {D.}~\bibnamefont
  {Reich}}, \bibinfo {author} {\bibfnamefont {G.}~\bibnamefont {Gualdi}}, \
  and\ \bibinfo {author} {\bibfnamefont {C.}~\bibnamefont {Koch}},\ }\href@noop
  {} {\bibfield  {journal} {\bibinfo  {journal} {Phys. Rev. Lett.}\ }\textbf
  {\bibinfo {volume} {111}},\ \bibinfo {pages} {200401} (\bibinfo {year}
  {2013})}\BibitemShut {NoStop}%
\bibitem [{\citenamefont {Dankert}\ \emph {et~al.}(2009)\citenamefont
  {Dankert}, \citenamefont {Cleve}, \citenamefont {Emerson},\ and\
  \citenamefont {Livine}}]{Dankert2009}%
  \BibitemOpen
  \bibfield  {author} {\bibinfo {author} {\bibfnamefont {C.}~\bibnamefont
  {Dankert}}, \bibinfo {author} {\bibfnamefont {R.}~\bibnamefont {Cleve}},
  \bibinfo {author} {\bibfnamefont {J.}~\bibnamefont {Emerson}}, \ and\
  \bibinfo {author} {\bibfnamefont {E.}~\bibnamefont {Livine}},\ }\href@noop {}
  {\bibfield  {journal} {\bibinfo  {journal} {Phys. Rev. A}\ }\textbf {\bibinfo
  {volume} {80}},\ \bibinfo {pages} {012304} (\bibinfo {year}
  {2009})}\BibitemShut {NoStop}%
\bibitem [{\citenamefont {Perron}(1907)}]{Perron1907}%
  \BibitemOpen
  \bibfield  {author} {\bibinfo {author} {\bibfnamefont {O.}~\bibnamefont
  {Perron}},\ }\href@noop {} {\bibfield  {journal} {\bibinfo  {journal} {Math.
  Ann.}\ }\textbf {\bibinfo {volume} {64}},\ \bibinfo {pages} {248} (\bibinfo
  {year} {1907})}\BibitemShut {NoStop}%
\bibitem [{\citenamefont {Frobenius}(1912)}]{Frobenius1912}%
  \BibitemOpen
  \bibfield  {author} {\bibinfo {author} {\bibfnamefont {G.}~\bibnamefont
  {Frobenius}},\ }\href@noop {} {\bibfield  {journal} {\bibinfo  {journal}
  {Sitzungsber. K\"onigl. Preuss. Akad. Wiss.}\ }\textbf {\bibinfo {volume}
  {-}},\ \bibinfo {pages} {456} (\bibinfo {year} {1912})}\BibitemShut {NoStop}%
\bibitem [{\citenamefont {Meyer}(2000)}]{Meyer2000}%
  \BibitemOpen
  \bibfield  {author} {\bibinfo {author} {\bibfnamefont {C.}~\bibnamefont
  {Meyer}},\ }\href@noop {} {\emph {\bibinfo {title} {Matrix analysis and
  applied linear algebra}}}\ (\bibinfo  {publisher} {SIAM},\ \bibinfo {year}
  {2000})\ p.\ \bibinfo {pages} {655}\BibitemShut {NoStop}%
\bibitem [{\citenamefont {Jordan}\ and\ \citenamefont
  {Wigner}(1993)}]{jordan1993paulische}%
  \BibitemOpen
  \bibfield  {author} {\bibinfo {author} {\bibfnamefont {P.}~\bibnamefont
  {Jordan}}\ and\ \bibinfo {author} {\bibfnamefont {E.~P.}\ \bibnamefont
  {Wigner}},\ }in\ \href@noop {} {\emph {\bibinfo {booktitle} {The Collected
  Works of Eugene Paul Wigner}}}\ (\bibinfo  {publisher} {Springer},\ \bibinfo
  {year} {1993})\ pp.\ \bibinfo {pages} {109--129}\BibitemShut {NoStop}%
\bibitem [{\citenamefont {Whitfield}\ \emph {et~al.}(2011)\citenamefont
  {Whitfield}, \citenamefont {Biamonte},\ and\ \citenamefont
  {Aspuru-Guzik}}]{whitfield2011simulation}%
  \BibitemOpen
  \bibfield  {author} {\bibinfo {author} {\bibfnamefont {J.~D.}\ \bibnamefont
  {Whitfield}}, \bibinfo {author} {\bibfnamefont {J.}~\bibnamefont {Biamonte}},
  \ and\ \bibinfo {author} {\bibfnamefont {A.}~\bibnamefont {Aspuru-Guzik}},\
  }\href@noop {} {\bibfield  {journal} {\bibinfo  {journal} {Molecular
  Physics}\ }\textbf {\bibinfo {volume} {109}},\ \bibinfo {pages} {735}
  (\bibinfo {year} {2011})}\BibitemShut {NoStop}%
\bibitem [{\citenamefont {Bravyi}\ and\ \citenamefont
  {Kitaev}(2002)}]{bravyi2002fermionic}%
  \BibitemOpen
  \bibfield  {author} {\bibinfo {author} {\bibfnamefont {S.~B.}\ \bibnamefont
  {Bravyi}}\ and\ \bibinfo {author} {\bibfnamefont {A.~Y.}\ \bibnamefont
  {Kitaev}},\ }\href@noop {} {\bibfield  {journal} {\bibinfo  {journal} {Annals
  of Physics}\ }\textbf {\bibinfo {volume} {298}},\ \bibinfo {pages} {210}
  (\bibinfo {year} {2002})}\BibitemShut {NoStop}%
\bibitem [{\citenamefont {Tranter}\ \emph {et~al.}(2015)\citenamefont
  {Tranter}, \citenamefont {Sofia}, \citenamefont {Seeley}, \citenamefont
  {Kaicher}, \citenamefont {McClean}, \citenamefont {Babbush}, \citenamefont
  {Coveney}, \citenamefont {Mintert},\ and\ \citenamefont
  {Love}}]{Tranter2015}%
  \BibitemOpen
  \bibfield  {author} {\bibinfo {author} {\bibfnamefont {A.}~\bibnamefont
  {Tranter}}, \bibinfo {author} {\bibfnamefont {S.}~\bibnamefont {Sofia}},
  \bibinfo {author} {\bibfnamefont {J.}~\bibnamefont {Seeley}}, \bibinfo
  {author} {\bibfnamefont {M.}~\bibnamefont {Kaicher}}, \bibinfo {author}
  {\bibfnamefont {J.}~\bibnamefont {McClean}}, \bibinfo {author} {\bibfnamefont
  {R.}~\bibnamefont {Babbush}}, \bibinfo {author} {\bibfnamefont
  {P.}~\bibnamefont {Coveney}}, \bibinfo {author} {\bibfnamefont
  {F.}~\bibnamefont {Mintert}}, \ and\ \bibinfo {author} {\bibfnamefont
  {F.~W.~P.}\ \bibnamefont {Love}},\ }\href@noop {} {\bibfield  {journal}
  {\bibinfo  {journal} {Int. J. Quant. Chem.}\ }\textbf {\bibinfo {volume}
  {115}},\ \bibinfo {pages} {1431} (\bibinfo {year} {2015})}\BibitemShut
  {NoStop}%
\bibitem [{\citenamefont {Lafarge}\ \emph {et~al.}(1993)\citenamefont
  {Lafarge}, \citenamefont {Joyez}, \citenamefont {Esteve}, \citenamefont
  {Urbina},\ and\ \citenamefont {Devoret}}]{Lafarge1993}%
  \BibitemOpen
  \bibfield  {author} {\bibinfo {author} {\bibfnamefont {P.}~\bibnamefont
  {Lafarge}}, \bibinfo {author} {\bibfnamefont {P.}~\bibnamefont {Joyez}},
  \bibinfo {author} {\bibfnamefont {D.}~\bibnamefont {Esteve}}, \bibinfo
  {author} {\bibfnamefont {C.}~\bibnamefont {Urbina}}, \ and\ \bibinfo {author}
  {\bibfnamefont {M.}~\bibnamefont {Devoret}},\ }\href@noop {} {\bibfield
  {journal} {\bibinfo  {journal} {Nature}\ }\textbf {\bibinfo {volume} {365}},\
  \bibinfo {pages} {422} (\bibinfo {year} {1993})}\BibitemShut {NoStop}%
\bibitem [{\citenamefont {Dallaire-Demers}\ and\ \citenamefont
  {Wilhelm}(2016{\natexlab{b}})}]{Dallaire-Demers2016}%
  \BibitemOpen
  \bibfield  {author} {\bibinfo {author} {\bibfnamefont {P.-L.}\ \bibnamefont
  {Dallaire-Demers}}\ and\ \bibinfo {author} {\bibfnamefont {F.}~\bibnamefont
  {Wilhelm}},\ }\href@noop {} {\bibfield  {journal} {\bibinfo  {journal} {Phys.
  Rev. A}\ }\textbf {\bibinfo {volume} {93}},\ \bibinfo {pages} {032303}
  (\bibinfo {year} {2016}{\natexlab{b}})}\BibitemShut {NoStop}%
\bibitem [{\citenamefont {Combes}\ \emph {et~al.}(2017)\citenamefont {Combes},
  \citenamefont {Granade}, \citenamefont {Ferrie},\ and\ \citenamefont
  {Flammia}}]{Combes2017}%
  \BibitemOpen
  \bibfield  {author} {\bibinfo {author} {\bibfnamefont {J.}~\bibnamefont
  {Combes}}, \bibinfo {author} {\bibfnamefont {C.}~\bibnamefont {Granade}},
  \bibinfo {author} {\bibfnamefont {C.}~\bibnamefont {Ferrie}}, \ and\ \bibinfo
  {author} {\bibfnamefont {S.}~\bibnamefont {Flammia}},\ }\href@noop {}
  {\enquote {\bibinfo {title} {Logical randomized benchmarking},}\ } (\bibinfo
  {year} {2017}),\ \bibinfo {note} {arXiv:1702.03688}\BibitemShut {NoStop}%
\end{thebibliography}%
\bibliographystyle{apsrev4-1}
\end{document}